\documentclass[onecollarge,natbib]{svjour2}
\bibpunct{[}{]}{;}{n}{}{,} 
\smartqed  
\usepackage{graphicx}
\usepackage{mathptmx}      
\usepackage{multirow}
\journalname{Few-Body Systems}
\begin{document}

\title{Universal behavior of few-boson systems using potential models}

\author{A. Kievsky, M. Viviani, R. \'Alvarez-Rodr\'\i guez, M. Gattobigio, A. Deltuva}

\institute{A. Kievsky \at Istituto Nazionale di Fisica Nucleare,
              Largo Pontecorvo 3, 56127 Pisa, Italy\\
              \email{kievsky@pi.infn.it}           
           \and
           M. Viviani \at
 Istituto Nazionale di Fisica Nucleare,
              Largo Pontecorvo 3, 56127 Pisa, Italy
           \and
           R. \'Alvarez-Rodr\'\i guez \at
Escuela T\'ecnica Superior de Arquitectura, Universidad
Polit\'ecnica de Madrid, Avda. Juan Herrera 4, E-28040 Madrid, Spain
           \and
           M. Gattobigio \at
Universit\'e de Nice-Sophia Antipolis, Institut Non-Lin\'eaire de
Nice,  CNRS, 1361 route des Lucioles, 06560 Valbonne, France
           \and
           A. Deltuva \at
Institute of Theoretical Physics and Astronomy, Vilnius University,
A. Go\u{s}tauto St. 12, LT-01108 Vilnius, Lithuania
}

\date{Received: date / Accepted: date}

\maketitle

\begin{abstract}
The universal behavior of a three-boson system close to the unitary limit
is encoded in a simple dependence of many observables in terms of few
parameters. For example the product of the three-body parameter $\kappa_*$ and the
two-body scattering length $a$, $\kappa_* a$ depends on the angle $\xi$
defined by $E_3/E_2=\tan^2\xi$. A similar dependence is observed in the
ratio $a_{AD}/a$ with $a_{AD}$ the boson-dimer scattering length.
We use a two-parameter potential to determine this simple behavior
and, as an application, to compute $a_{AD}$ for the case of three $^4$He atoms.
\end{abstract}

\section{Introduction}
\label{intro}
The study of few-boson systems close to the unitary limit is an intense
subject of research nowadays. For identical bosons, the unitary limit is 
defined when the two-body scattering 
length $a\rightarrow\infty$. In this limit the three-boson system shows the
Efimov effect~\cite{efimov1,efimov2}. Moreover, close to the unitary limit
the system manifests universal behavior: the details of the two-body
interaction are not important and its spectrum is determined essentially by $a$ and the 
three-body parameter $\kappa_*$ which defines the energy $E_*=\hbar^2\kappa^2_*/m$
of level $n_*$ at the unitary limit, here $m$ is the boson mass (for a recent review see 
Ref.~\cite{report}). At the unitary limit the spectrum shows a
discrete scaling invariance (DSI): an infinite series of bound states appears 
distributed geometrically and accumulates at zero energy. The ratio of two 
consecutive energy states is constant,
$E^n_3/E^{n+1}_3=e^{2\pi/s_0}$, with the universal number $s_0\approx 1.00624$.

The universal characteristics of the system  can be exploited studying the dynamics using potential
models constructed in such a way that the control parameters of the spectrum are
reproduced. For example a two-parameter potential as a gaussian can be used to this
aim~\cite{kievsky2015,kievsky2016}. Universal behavior manifests in a simple dependence
of many observables on the angle defined by the ratio $E_3/E_2=\tan^2\xi$. This is the
case of the product $\kappa_*a$ and, in the case in which the two bosons forms a dimer,
the ratio $a_{AD}/a$ between the boson-dimer scattering length $a_{AD}$ and the two-body
scattering length $a$. These relations are exactly fulfilled in the zero-range limit and,
as we will show, range corrections can be introduced using the potential models.
In fact the gaussian potentials can be used to determine the dependence on the
angle $\xi$ of the observables defined above. The first one, $\kappa_*a$, is used to define
the gaussian level function~\cite{kievsky2015}. Moreover a gaussian potential verifying 
$E_3/E_2=\tan^2\xi$ can be used to compute the ratio $a_{AD}/a$ at that angle.
After introducing range corrections we compute this ratio and, as an application, we determine
$a_{AD}$ for the case of a system composed by three $^4$He atoms. 

\section{The gaussian level function}
\label{sec:1}

In the case of a zero-range interaction the $L=0$ spectrum of three bosons 
is determined by the Efimov radial law
  \begin{eqnarray}
    \label{eq:energyzrA}
      E_3^n/(\hbar^2/m a^2) = \tan^2\xi_n , \\
      \kappa_*a = {e}^{(n-n^*)\pi/s_0} 
      \frac{{e}^{-\Delta(\xi)/2s_0}}{\cos\xi}\, .
    \label{eq:energyzrB}
  \end{eqnarray}
    \label{eq:energyzr}
where $a$ is the two-body scattering length and $E_3^n$ is the energy of
level $n$. The binding momentum $\kappa_*$ gives the energy of the system at the unitary
limit. The spectrum is determined by the knowledge of the universal
function $\Delta(\xi)$ which is equal for all levels (a parametrization
of the universal function is given in ~\cite{report}). In fact knowing one energy value
the complete spectrum is determined. The zero-range theory is not always sufficiently
accurate to describe real systems and range corrections have to be introduced at some
level.
Since close to the unitary limit the details of the interaction are not important,
it is possible to construct a potential model in order to
capture the essential ingredients of the dynamics. Following this strategy a minimal
information that preserves universal behavior is encoded in the effective range
expansion for two particles, $k\cot\delta=-1/a + r_{\rm eff}k^2/2$. With $k$ the
relative momentum and $r_{\rm eff}$ the effective range. At low energies this 
perturbative expansion is well fulfilled and, in the case of shallow states,
it can be extended to negative energies relating $a$, $r_{\rm eff}$ and the
two-body energy $E_2$:
\begin{equation}
\frac{1}{a_B} = \frac{1}{a}+\frac{r_{\rm eff}}{2a_B^2} \;\ ,
\end{equation}
where we have introduced the energy length from the relation $E_2=\hbar^2/ma_B^2$.
Therefore a two-parameter potential describing $E_2$ and $a$ will also describe
$r_{\rm eff}$. Accordingly we define a local and a non local gaussian potential 
\begin{eqnarray}
& V&^L(r)=- V^L_0 e^{-r^2/r_0^2} \;\; ,
\label{eq:pot1} \\
& V&^{NL}(k,k')=- V^{NL}_0 e^{-k^2/k_0^2} e^{-{k'}^2/k_0^2} \;\; ,
\label{eq:pot2}
\end{eqnarray}
with the strengths $V^L_0$, $V^{NL}_0$ and the ranges $r_0$, $k_0^{-1}$
determined to describe particular values of $a$ and $E_2$ of a two-boson system. 
If these values are experimental values we call this set of values a physical point.
Once this point is fixed the strength of
the potential can be varied to reach the unitary limit.
With the potentials defined above the lengths, momenta and energies scale with
$r_0$, $k_0$ and $\hbar^2/mr_0^2$ (or $\hbar^2 k_0^2/m$), respectively. Therefore
the local gaussian (LG) and the nonlocal gaussian (NLG) potentials define a particular path to 
the unitary limit that encompasses all the local and nonlocal gassian potentials. 
In particular, for the ground state,
the values of the effective range and strength at unitary are 
$r_{\rm eff}=1.43522r_0$ and $\lambda V^L_0=2.6840 \hbar^2/mr_0^2$ (local gaussian)
and $r_{\rm eff}=3.19154/k_0$ and $\lambda V^L_0=0.126987 \hbar^2 k^2_0/m$ (nonlocal gaussian).

The potentials defined in Eqs.(\ref{eq:pot1},\ref{eq:pot2}) can be used to describe the
three-boson system close to the unitary limit. The Efimov law of Eq.(\ref{eq:energyzr})
suggests the following representation of the gaussian $L=0$ spectrum of three-bosons 
  \begin{eqnarray}
    \label{eq:energyfrA}
      E_3^n/E_2= \tan^2\xi_n , \\
      \kappa^n_*a_B = \frac{{e}^{-\widetilde\Delta_n(\xi)/2s_0}}{\cos\xi}\, .
    \label{eq:energyfrB}
  \end{eqnarray}
    \label{eq:energyfr}
where $\kappa^n_*$ defines the energy of level $n$ at the unitary limit,
$E_*^n=\hbar^2(\kappa_*^n)^2/m$, and the gaussian level function is defined as
\begin{equation}
\widetilde\Delta_n=s_0\ln\left(\frac{E_3^n+E_2}{E_*^n}\right)\;\;\; .
\end{equation}
The scaling properties of the gaussian potentials are such that the level function
is the same for all local gaussian and for all nonlocal gaussian, being the local
and nonlocal level functions slightly different for the ground state level $n=0$.
As $n>0$ both level functions tend to be equal and tend to the zero-range function, 
$\widetilde\Delta_n\rightarrow \Delta$. This behavior is show in Fig.1 in which the
LG and NLG level functions $\widetilde\Delta_n(\xi)$ are shown for the ground and
first excited state levels $n=0,1$. The
universal zero-range function is shown as well and completely overlap with
the NLG level function calculated for the third excited state $\widetilde\Delta_3(\xi)$.
In particular the ground state level function $\widetilde\Delta_0(\xi)$ incorporate range corrections and
can be used to determine the corresponding three-body parameter $\kappa_*^0$~\cite{kievsky2015,kievsky2016}.

\begin{figure}[h]
\vspace{1.5cm}
    \begin{center}
      \includegraphics[width=\linewidth]{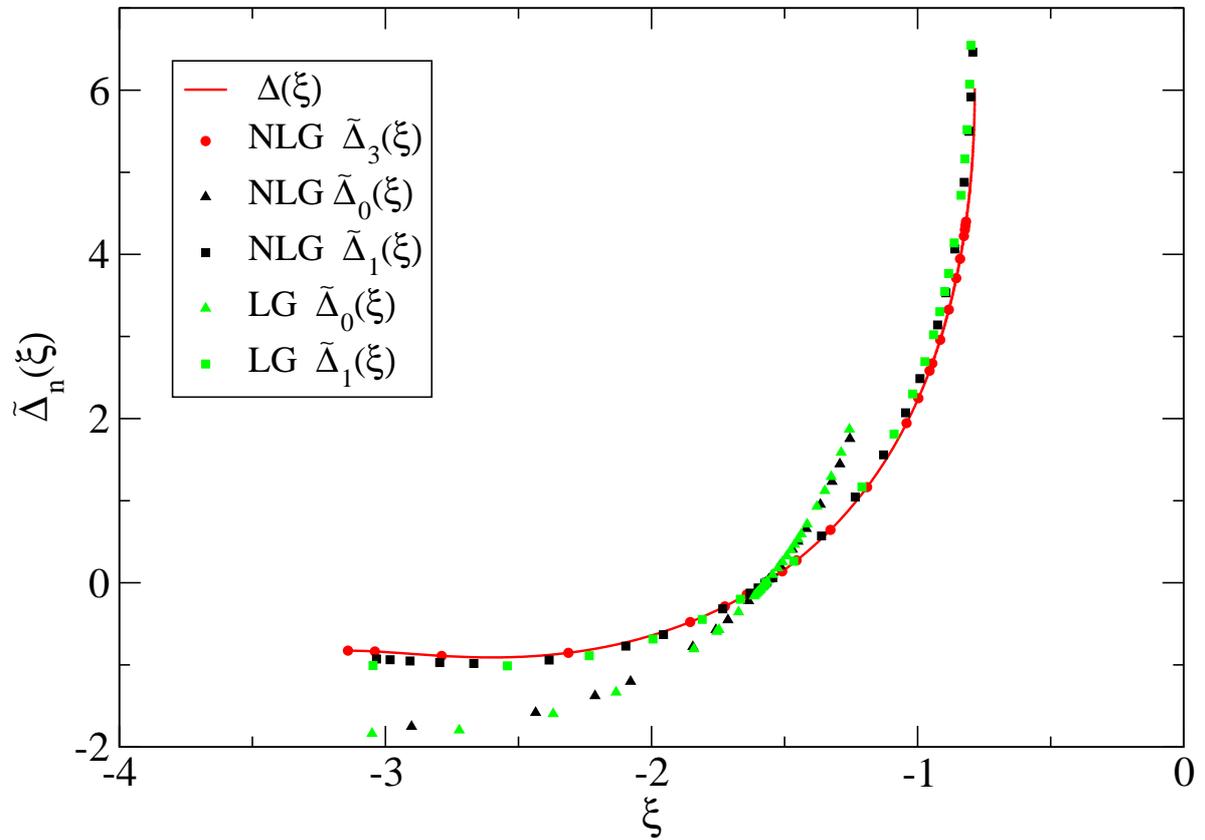}
    \end{center}
    \caption{The LG and NLG level functions $\widetilde\Delta_n(\xi)$
  are shown for the ground state and first excited state levels $n=0,1$.
  In the NLG case the third excited state ($n=3$) level function $\widetilde\Delta_3(\xi)$
  (red circles) completely overlap with the zero-range universal function $\Delta(\xi)$ 
  (red solid curve).
     }
\label{fig:fig1}
\end{figure}

We now discuss the $\xi$ dependence of the ratio $a_{AD}/a$.
As discussed by Efimov~\cite{efimov3} (see also Ref.~\cite{report}), in the
zero-range limit, the ratio $a_{AD}/a$ has the form
\begin{equation}
\frac{a_{AD}}{a}=d_1+d_2\tan[s_0\ln(\kappa_*a)+d_3]
\end{equation}
with $d_1$, $d_2$ and $d_3$ universal numbers (numerical values are given in Refs.~\cite{report,kievsky2013}). 
Due to the $\xi$ dependence of
the product $\kappa_* a$, the above ratio depends on this angle too.
Accordingly it is possible to study this ratio using potentials models.
Following Ref.~\cite{kievsky2013} we introduce range corrections by studying the
ratio $a_{AD}/a_B$. We calculate it for different values of $\xi$ and, in 
particular, we consider valid the following relation at the same $\xi$ value:
\begin{equation}
\frac{a_{AD}}{a_B}=\frac{a_{AD}}{a_B}\mid_{gaussian},
\end{equation}
where the first ratio refers to the values predicted by any realistic interaction or
experimental values whereas $gaussian$ refers to the values calculated with the LG or NLG potentials. 
As an application we calculate the value of the atom-dimer scattering length $a_{AD}$
for a system composed by three $^4$He atoms. This system is well described by the realistic 
LM2M2 potential of Aziz~\cite{aziz} for which $E_2=1.303\;$mK. Using this potential 
the three-boson ground state energy is $E^0_3=126\;$mK and, accordingly, the angle $\xi$ 
is obtained from $E_3/E_2=\tan^2\xi=97.0$
This ratio can be reproduced using a gaussian potential and with that potential
the value of $a_{AD}$ can be calculated using standard methods. After this
straightforward procedure we obtain 
\begin{equation}
a_{AD}|_{LM2M2}=a_B|_{LM2M2}\frac{a_{AD}}{a_B}|_{gaussian}\approx 212 {\rm a_0}
\end{equation}
to be compared to $217\;$a$_0$ obtained from a direct calculation using the LM2M2
interaction~\cite{kolga2009,kievsky2011}.
As can we see the simple dependence of the ratio $a_{AD}/a_B$ on the angle $\xi$ 
is well fulfilled and, furthermore, the gaussian description reproduces 
this dependence with good approximation.

\section{Conclusions}
\label{sec:2}
We have studied the three-boson system using potential models of the gaussian form.
This simple representation of the interaction seems to capture the main aspects of 
the low energy dynamics close to the unitary limit.
We have looked at the ratio $E_3/E_2$ defining the angle $\xi$. At the unitary limit 
it corresponds to $\xi=-\pi/2$. The DSI manifests at fixed values of $\xi$, this means that
the infinite series of states observed at $\xi=-\pi/2$ can be observed 
also along the line in which $\xi$ remains constant and the energy ratios along that
line verify the same relation $E^n_3/E^{n+1}_3=e^{2\pi/s_0}$. It should be noticed that the case
$\xi=-\pi/2$ corresponds to a single value of $E_2$ (equal to zero), 
whereas $E_2$ varies along the line in which $\xi$ is constant and therefore
the values of $E_2$ at which $E^n_3$ and $E^{n+1}_3$ are calculate are different.
This property allows to study DSI at finite values of $E_2$.

In the present work we have exploited the property that some observables close to the
unitary limit have a simple dependence, they are functions of the angle $\xi$. First we 
have analysed the product $\kappa_*^n a_B$ from which we have defined the gaussian level
function $\widetilde\Delta_n(\xi)$. In second place we have analysed the ratio
$a_{AD}/a_B$ and the gaussian values for this ratio have been used to determine 
$a_{AD}$ for the case of atom-dimer collision of three $^4$He atoms at zero energy.
From the results we observe that $\xi$ dependence is well verified
and, the gaussian model reproduces well the LM2M2 value for $a_{AD}$.
Further investigations along this line are under way.

 \begin{acknowledgements}
This work was partly supported by Ministerio de Econom\'{i}a y Competitividad
(Spain) under contracts MTM2015-63914-P and FPA2015-65035-P. Part of the
calculations of this work were performed in the high capacity cluster for
Physics, funded in part by Universidad Complutense de Madrid and in part
with Feder funds as a contribution to the Campus of International
Excellence of Moncloa, CEI Moncloa. R.A.R thanks Ministerio de
Educaci\'on, Cultura y Deporte (Spain) for the “Jos\'e Castillejo” fellowship
in the framework of Plan Estatal de Investigaci\'on Cient\'{i}fica y T\'ecnica y
de Innovaci\'on 2013-2016.
 \end{acknowledgements}

\end{document}